# Atomic bonding and Electronic binding energy of two-dimensional Bi/Li(110) heterojunctions via BOLS-BB model

Maolin Bo*, Liangjing Ge, Jibiao Li, Lei Li, Chuang Yao, Zhongkai Huang

Key Laboratory of Extraordinary Bond Engineering and Advanced Materials Technology (EBEAM) Chongqing, Yangtze Normal University, Chongqing 408100, China

**Corresponding Author:** *E-mail: bmlwd@yznu.edu.cn (Maolin Bo)

**ABSTRACT**

Combining the bond-order-length-strength (BOLS) and atomic bonding and electronic model (BB model) with density functional theory (DFT) calculations, we studied the atomic bonding and electronic binding energy behavior of Bi atoms adsorbed on the Li(110) surface. We found that the Bi atoms adsorbed on the Li(110) surface form two-dimensional (2D) geometric structures, including letter-, hexagon-, galaxy-, crown-, field-, and cobweb-shaped structures. Thus, we obtained the following quantitative information: (i) the field-shaped structure can be considered the bulk structure; (ii) the field-shaped structure of Bi atom formation has a 5$d$ energy level of 22.727 eV, and in the letter shape structure, this energy is shifted to values greater than 0.342 eV;and (iii) the Bi/Li(110) heterojunction transfers charge from the inner Li atomic layer to the outermost Bi atomic layer. In addition, we analyzed the bonding and electronic dynamics involved in the formation of the Bi/Li(110) heterojunctions using residual density of states. This work provides a theoretical reference for the fine tuning of binding energies and chemical bonding at the interfaces of 2D metallic materials.

INTRODUCTION

With developments in nanotechnology and nanomanufacturing technologies, it has become possible to isolate synthesized materials with one (or few) atomic thickness[1-4]. Graphene was the first atomic thickness material, having been separated from graphite by Geim and Novoselov[1]. This material is called a 2D material because electrons can move only in two directions (in-plane), being confined in the third direction. Thereafter, several layered 2D materials have risen to prominence, including transition metal carbon dihalides[5], hexagonal boron nitride[6], and black phosphorus[7]. The confinement of this system provides graphene and other such materials with excellent electrical and optical properties, and as such they can be widely used in nanoelectronics and energy research[8-14]. Prior to the discovery of 2D semiconductor materials, the interfaces of 2D metal heterojunctions were considered to be very complex and poorly defined, rendering it difficult to quantify the active junction area and the physical properties associated with the junction[15].

In fact, a metal heterojunction is easy to form when a metal material is grown on a suitable substrate[16,17]. Furthermore, 2D metal material growth requires the use of a suitable substrate material. Artificially designed and synthesized 2D metal heterojunctions, compared with bulk materials, have many peculiar physical and chemical properties[18-23]. For example, 2D metal Bi/SiC hexagon structures heterojunctions have an indirect gap of 0.67 eV[24]. However, A heavy atom like Bi multiple bonding patterns are possible. we calculated multiple bonding patterns Bi atoms on the Li(110) surface. we found that there is agglomeration of Bi atoms on the interface. The agglomeration of Bi atoms on the Li(110) interface form the 2D heterojunction. In addition, the electronic properties of 2D metal heterojunctions can be modulated as a function of an applied gate voltage, leading to fundamentally new device phenomena and providing opportunities for the tuning of device properties[25,26].

In this study, we calculated the energies of 2D metal Bi/Li(110) heterojunctions using first-principles methods. We found that Bi atoms were adsorbed on the surface of Li(110), forming 2D geometric structures in the shapes of a letter, hexagon, galaxy, crown, field, and cobweb. Unlike the magnetic properties of the 2D Metal Sc/Li (110) heterojunctions study, [27] we used the BOLS-BB model to study the atomic bonding and electronic binding of Bi/Li (110) heterojunctions. The bonding and electron dynamics processes at the Bi/Li(110) heterojunction were studied, and parameter information about the related bonding and electronic structure were obtained. By establishing the concept of the combined BOLS–BB model, the internal mechanism of the electronic binding energy shift and chemical bond relaxation at Bi/Li(110) metal heterojunctions is revealed. This work provides a new approach for the calculation of the atomic bonding and electronic binding energies of 2D metal heterojunctions.

**RESULTS AND DISCUSSION**

We used first-principles methods to calculate Bi/Li(110) heterostructures, which are obtained by adsorption of Bi atoms at different positions on the Li(110) surface. The optimized geometry of Bi/Li(110) heterostructures is shown in **Fig. 2**. We found that Bi atoms adsorbed on the Li(110) surface formed various 2D geometric structures, namely, letter, hexagon, galaxy, crown, field, and cobweb shapes. This is similar to the 2D structure formed by the adsorption of Sc[27] and Y[28] atoms on the Li(110) surface. In addition, the height $h$ between the Bi and Li(110) layers is the distance between the Bi atomic layer and Li(110) atomic layer in the direction of the thickness of the slab. The calculated interlayer heights $h_x$ of these structures are 0.95, 2.09, 1.10, 1.59, 2.36, and 1.76 Å, respectively, as shown in **Table 1**.

We calculated the total energy of Bi and Li(110) and that of Bi/Li(110) in the letter-, hexagon-, galaxy-, crown-, field-, and cobweb-shaped structures. Thus, the formation energy $E_{form}$ was calculated as the energy difference between these values, using the equation[29-31] $E_{form} = E_{Bi/Li} - E_{Bi} - E_{Li}$ ; where $E_{Bi/Li}$, $E_{Bi}$, and $E_{Li}$ are the total energies of the relaxed Bi/Li(110) heterostructure, isolated Bi, and Li monolayers, respectively. these values are presented in **Table 1**. The formation energies of the letter-, hexagon-, galaxy-, crown,- field-, and cobweb-shaped heterojunctions are -6.77, -7.87, -9.78, -7.98, -7.83, and -7.88 eV, respectively. Therefore, we confirm that both the Bi/Li(110) heterojunctions are stable geometric structures.

The work function is the minimum energy needed to remove an electron from a solid to a point immediately outside the solid surface or the energy needed to move an electron from the Fermi energy level to the vacuum. The value of the work function is an indication of the strength of the BE of the electrons in the metal. The larger the work function, the less likely it is for the electrons to leave the metal. We calculated the work functions of Bi, Li(110), and Bi/Li(110) heterojunctions, as shown in **Table 1**. For the same structure, we found that the work functions are in the order Bi > Bi/Li(110) > Li(110). The results show that Bi/Li(110) heterojunctions effectively reduce the work function of Bi metal.

**Fig. 3** shows the density of states(DOS) diagram for the Bi 5*d* and Li 1*s* orbitals of the letter-, hexagon-, galaxy-, crown-, field-, and cobweb-shaped structures calculated by DFT. **Fig. 3a** shows that the Bi 5*d* orbital has negative BE shifts. Comparing the data in **Fig. 3 and Table 2**, we observe that the Bi 5*d* orbital binding energies of the letter-,

hexagonal-, galaxy-, crown-, field-, and cobweb-shaped structures are 22.385, 22.491, 22.669, 22.671, 22.727, and 22.521 eV, and the BE of the structures increases in the order field shape > crown shape > galaxy shape > cobweb shape > hexagon shape > letter shape.

To calculate the change in the interface bonding at Bi/Li(110) heterojunctions caused by Bi metal doping, we should obtain the interface BE shift $\Delta E_v(i) = E_v(x) - E_v(B)$ and the energy level width of the bulk $\Delta E_{5d}(w_B) = 1.13 \text{ eV}$. For the value of $E_{5d}(B)$, we used the peak value of the BE of the field-shaped structure as a reference. Using **Eq. 13, 14 and 15**, we calculated the RBED $\delta E_d$, RLBS $\delta\varepsilon_x$, and RBER $\delta\gamma_x$ of the letter-, hexagon-, galaxy-, crown-, field-, and cobweb-shaped structures, as shown in **Table 1 and Fig. 4**. By calculating the atomic bond parameters, we found that compared to the field-shaped structure, the letter-, hexagon-, galaxy-, crown-, and cobweb-shaped structures have negative values of the RBED $\delta E_d$ and RBER $\delta\gamma_x$ and positive values of the RLBS $\delta\varepsilon_x$. This indicates the properties of these shaped structures compared to those of a layer of Bi atoms adsorbed on the Li(110) surface.

Using DFT calculations to calculate the energies and structures of Bi/Li(110) heterojunctions, we found that electron transfer occurs within the valence electron band after Bi is adsorbed on the Li(110) surface. The residual **DOS (RDOS)** results are obtained as the difference between spectra collected from a surface before and after it is physically or chemically conditioned[32]. From the RDOS data, we can obtain the distribution of the electronic structure and bonding near the valence band at the Fermi level ($E_F$). **Fig. 5** shows the RDOS data for the letter-, hexagon-, galaxy-, crown-, field-, and cobweb-shaped structures. Bi atoms are adsorbed on the Li(110) surface (**Fig. 2**). Consistent with the prediction of the BBB theory[33], there are four characteristic regions of the DOS plot: Bi–Li bonding states (from -4 to -6 eV), nonbonding states (from -2 to -4 eV), electron ($Li^{\delta-}$)–hole ($Bi^{\delta+}$) (-1 eV and 1 eV), and antibonding states (from +2 to +4 eV). For the valence band, charge is transferred from a lower energy level to a higher energy level, and the atomic polarization is an important parameter for electrons and holes.

Regarding the deformation charge density values (**Table 2 and Fig. 6**), it should be noted that a positive sign indicates that electrons are accepted, and the charge is transferred from Li atoms to Bi atoms. **Figs. 6** and **7** present the electron dynamics of the bonding at the heterojunctions. The low coordination $Bi^{\delta+}$ atom on the surface causes the $Li^{\delta-}$ atom to produce an electronic nonbonding state. The nonbonding lone pair of electrons polarizes the $Bi^{\delta+}$ electrons to form $Bi^{\delta+}$–$Li^{\delta-}$ dipole moments. Simultaneously, the polarization appears as Bi $5d$ core bands. The electron BE of the core band has a negative shift, as shown in **Fig. 7**. The characteristic electron ($Li^{\delta-}$)–hole ($Bi^{\delta+}$) bands in the DOS, at -1.0 and 1.0 eV, are produced by the $Bi^{\delta+}$–$Li^{\delta-}$ dipole moment, which produces a bond between the heteroatoms that is stronger than either the Li–Li or Bi–Bi bonds. The stronger electronic interaction reduces the energy of the surface work function, and the electron BE of the Bi $5d$ core band balances the electron distribution at the interface. The antibonding state (+2 to +4 eV) arises as a result of the polarization of Bi electrons by the isolated $Bi^{\delta+}$–$Li^{\delta-}$ dipole moment. In addition, it should be emphasized that the RDOS results not only provide an effective numerical calculation method for quantitatively studying atomic bonding and electronic behavior at Bi/Li(110) heterojunctions, but it also provides a new method for the formation of Bi/Li(110) heterostructures.

Unlike the valence band, in which charge mixing and atomic *s*-, *p*-, and *d*- orbital mixing occur after the formation of the heterojunctions, the electron transfer in the core bands of the atomic energy levels in the DOS distribution occurs as a result of the effects of potential energy or bonding state. Because for weak hybridization (polarization) the bonding state is ligand-like (Bi atoms), and the antibonding state mostly metal-like (Li atoms). Our calculations revealed that the Millikan charge of the first layer of Li atoms is positive, indicating that electrons are lost. The negative charge of the Bi atoms implies that electrons are obtained. The electronic structure of the Bi/Li(110) heterojunctions calculated here is similar to that of the PN junction structure of the semiconductor, as shown in **Fig. 7a**.

We calculated the potential energy functions of the surfaces as a function of the positive and negative charges, $V_{surface}(r_i) = (\frac{1}{4\pi\varepsilon_0})\frac{q_1(+)q_2(-)}{r_i}$. The calculated potential energies of the surfaces are all negative, indicating that chemical bonds are formed between the Li and Bi atoms. The more negative the potential energy, the better the interface bonding performance, and the formation of the PN junction structure will strengthen the bonding performance of the interface atoms. As both Bi and Li are metals, they are not the same as semiconductors, and thus they have no band gap. Therefore, here we present the energy band structure of the core band, as shown in **Fig. 7b**. The PN junction structure of the metal heterojunctions allows a part of the excess electron density from the Li atoms to be transferred to different energy levels of the Bi atoms. The filling of different Bi atomic energy levels causes a negative shift in the BE of electrons. Therefore, the formation of the PN junction structure of the metal heterojunctions will cause a shift in the electron BE.

**CONCLUSIONS**

We investigated the geometric, electronic, energetic, and bonding properties of Bi/Li(110) heterojunctions. . The results are summarized below:

1) The formation of clusters on the surface requires a large number of atoms, and we calculate that the number of atoms is less than 10 atoms. Our results show that a small number of Bi atoms can also agglomerate on the surface to form a 2D heterostructure.

2) We found that the interface of the Bi/Li(110) heterojunction will form a PN junction. The results show that the PN junction formed by the metal heterojunction will result in a core level shift.

3) We used a combined BOLS–BB model to calculate parameter information for the atomic bonding and electronic properties of the Bi/Li(110) heterojunctions. The results of the BOLS–BB model provide an effective numerical calculation for studying the quantitative information about the bonding and electronic behavior of Bi/Li(110) heterostructures.

**Principles and methods**

**Tight binding approximation.** In the Tight binding approximation, the Hamiltonian is given by[34]

$$H = \left[ -\frac{\hbar^2 \nabla^2}{2m} + V_{atom}(r) \right] + V_{cry}(r)(1+\Delta_H)$$
$$= -\frac{\hbar^2}{2m}(\nabla - \frac{q}{\hbar c}A(\vec{r})i)^2$$

(1)

where $A(\vec{r}) \propto 1/\vec{r}$ is the vector potential, $m$ is electron mass, $\hbar$ is the Planck's constant, $c$ is speed of light and $q$ is the electron charge. The electronic binding energy (BE) of the $v$th energy band $E_v(x)$ is

$$E_v(0) = \left\langle v,i \left| -\frac{\hbar^2 \nabla^2}{2m} + V_{atom}(r) \right| v,i \right\rangle \qquad \text{(atomic level)}$$

(2)

$$E_v(B) - E_v(0) = \left\langle v,i | V_{crys}(r) | v,i \right\rangle + \sum_j f(k) \left\langle v,i | V_{crys}(r) | v,j \right\rangle$$

$$= \alpha_v (1 + \sum_j \frac{f(k) \cdot \beta_v}{\alpha_v}) \cong \alpha_v \qquad \text{(binding energy shift of bulk)}$$

(3)

$$\alpha_v = <v,i | V_{crys}(r) | v,i>; \quad \beta_v = <v,i | V_{crys}(r) | v,j>$$

(4)

$$\frac{E_v(x) - E_v(0)}{E_v(B) - E_v(0)} = \frac{\alpha(x)}{\alpha(B)}$$

. (5)

$$V_l = \int V_{crys}(r) e^{ikr} dr \cong \left\langle v,i | V_{crys}(r) | v,i \right\rangle$$

(6)

$$E_G = 2|V_l| = 2\left|\left\langle v,i | V_{crys}(r) | v,i \right\rangle\right|$$

(7)

In this definition, $E_v(B)$ and $E_v(0)$ are the energy levels of bulk atoms and an isolated atom, respectively; $\alpha_v$ is the exchange integral; and $\beta_v$ is the overlap integral, which contributes to the width of the energy band. $|v,i\rangle$ represents the wave function, with a periodic factor $f(k)$ in the form of $e^{-ikr}$, while $k$ is the wave vector. $V_{atom}(r)$ is the intra-atomic potential of the atom, $V_{crys}(r)$ is the potential of the crystal, and the interaction potential changes with the coordination environment and during chemical reactions. In the localized band of core levels, $\beta_v$ is very small, so $\alpha_v$ determines the energy shift of the core levels. $E_G$ is band gap.

**BOLS–BB model.** In the BOLS–BB model, the bond energy uniquely determines the impurity-induced core-level BE shift[20]:

$$z_x = \frac{12}{\left\{8\ln\left(\frac{2\Delta E'_v(x) - \Delta E_v(B)}{\Delta E_v(B)}\right) + 1\right\}} \quad (\Delta E_v(x) \geq 0)$$

(8)

$$\frac{E_v(x) - E_v(0)}{E_v(B) - E_v(0)} = \frac{\Delta E'_v(x) + \Delta E_v(B)}{\Delta E_v(B)} = \gamma = 1 + \Delta_H$$

(9)

where $z_x$ is the atomic coordination number of an atom in the $x$th atomic layer from the surface. The energy level of an isolated atom $E_v(0)$ is the unique reference energy, from which the electronic BE of the core energy levels are considered. The bulk component $E_v(B)$ is obtained from X-ray photoelectron spectroscopy (XPS) results and DFT calculations.

$$\vec{E}_x = -\frac{1}{2m}(\nabla - qA(\vec{r}_x)i)^2 \propto \frac{1}{2m}(p_x - \frac{q}{\vec{r}_x})^2; \quad (p_x = -i\hbar\nabla, \hbar = 1, c = 1)$$

(10)

In this definition, $p_x$ is the momentum, $r_x$ is the electron radius, and $z_{xb} = z_x/z_b$ is the atomic fractional coordination number.

$$\frac{\Delta E_v(w_x)}{\Delta E_v(w_B)} = \frac{z_x E_x}{z_b E_b} = z_{xb} c_x^{-m} = z_{xb} (\frac{d_x}{d_b})^{-m}$$

(11)

$$\frac{\Delta E_v(w_x)}{\Delta E_v(w_B)} \propto \frac{E_x}{E_b} = \gamma = 1 + \Delta_H = c_x^{-m} = (\frac{d_x}{d_b})^{-m}$$

(12)

$$\frac{\Delta E_v(w_x)}{\Delta E_v(w_B)} \cong \frac{\Delta E_v'(x) + \Delta E_v(B)}{\Delta E_v(B)} = \gamma$$

(13)

$$\Delta V_{cry}(r) = V_{cry}(r)(1 + \Delta_H) = \gamma V_{cry}(r) = (Z+1)\frac{1}{4\pi\varepsilon_0}\sum_i \langle v,i|\frac{e^2}{r_i}|v,i\rangle$$

(14)

$$E_v(x) - E_v(0) \approx -\langle v,i|V_{cry}(r)(1 + \Delta_H)|v,i\rangle = -(Z+1)\frac{1}{4\pi\varepsilon_0}\sum_i \langle v,i|\frac{e^2}{r_i}|v,i\rangle$$

(15)

$Z$ is the initial atom charge (neutral $Z = -1$(isolated atom), positively $Z = 0$ (bulk atoms), positively $Z = +|\delta\gamma|(\delta\gamma > 0)$ (charged positive atoms) and negatively $Z =$

$-|\delta\gamma|(\delta\gamma < 0)$ (charged negative atoms). Thus the core-electron BE shifts will be 0,

$-\sum_i \langle v,i| \frac{(1+|\delta\gamma|)}{4\pi\varepsilon_0} \frac{e^2}{r_i} |v,i\rangle, -\sum_i \langle v,i| \frac{1}{4\pi\varepsilon_0} \frac{e^2}{r_i} |v,i\rangle, -\sum_i \langle v,i| \frac{(1-|\delta\gamma|)}{4\pi\varepsilon_0} \frac{e^2}{r_i} |v,i\rangle$ and initially neutral of isolated atom, bulk atoms, singly charged positive and negative atoms, respectively.[18] **Eq. 12** provides estimates for the bond energy $E_x$, bond length $d_x$ and $\Delta E_v(w_B) \cong \Delta E_v(B)$ is the spectral full width of the bulk component ($w_B$) of the $v$th energy level; The width of the BE shift for the surface component ($w_x$) of the $v$th energy level is $\Delta E_v(w_x) = \Delta E_v(w_B) + \Delta E'_v(x)$; actual spectral intensities and shapes, however, are subject to polarization effects and measurement artifacts. We can calculate the chemisorption and defect-induced interface bond energy ratio $\gamma$ with the known reference value of $\Delta E'_v(x) = E_v(x) - E_v(B)$, $\Delta E_v(B) = E_v(B) - E_v(0)$ and $\Delta E_v(x) = E_v(x) - E_v(0)$ derived from the surface via DFT calculations and XPS analysis. Hence, we obtain

$$\delta\gamma = \frac{\Delta E_v(w_B) + \Delta E'_v(x)}{\Delta E_v(w_B)} - 1 = \gamma - 1; \quad \text{(RBER)}$$

(16)

$$\delta\varepsilon_x = d_x/d_b - 1 = \gamma^{-1} - 1 \quad \text{(RLBS)}$$

(17)

$$\delta E_d = (E_i/d_i^3)/(E_b/d_b^3) - 1 = \gamma^4 - 1 \quad \text{(RBED)}$$

(18)

$$\Delta E'_v(x) \approx -\delta\gamma \langle v,i|V_{cry}(r)|v,i\rangle \cong \beta_v = -\langle v,i|V_{cry}(r)|v,j\rangle \propto E_x - E_b$$

(19)

Thus, one can drive the interface relative bond energy ratio (RBER) parameter $\delta\gamma$ and elucidate via XPS analysis whether the bond energy strength determines the interface performance. If $\delta\gamma < 0$, the bond energy $E_x$ is reduced and the bond is weakened. Conversely, if $\delta\gamma > 0$, the bond energy increases and the bond becomes stronger. The relative local bond strain (RLBS) $\delta\varepsilon_x$ indicates the relative contraction of the atomic bond length $d_x$. The relative bond energy density (RBED) $\delta E_d$ is the energy density of the atomic bond with energy $E_i$.

**DFT calculations.** We calculated the geometric structure, atomic bonding, charge transfer, BE shifts, and electronic distribution of Bi/Li (110) heterojunctions by using DFT calculations. The optimal geometric configurations are shown in **Fig. 2**. The Vienna *ab initio* simulation package and plane-wave pseudopotentials are used in the calculations[36,37]. We also employed the Perde–Burke–Ernzerhof exchange-correlation potentials[38]; the plane-wave cutoff was 400 eV, and the vacuum space was 18 Å. The Brillouin zone was calculated with special *k*-points generated in an $8 \times 8 \times 1$ mesh grid. In the calculations, the energy converged to $10^{-6}$ eV and the force on each atom converged to <0.01 eV/Å. To consider long-range van der Waals interaction, DFT-D3 calculation was used.

**Acknowledgment:**

Financial support was provided by the National Natural Science Foundation of China (11947021 and 11904033), the Advanced Research Projects of Chongqing Municipal Science and Technology Commission (cstc2019jcyj-msxmX0674), and the Research Program of Chongqing Municipal Education Commission (Grant No. KJQN201901421).

**AUTHOR INFORMATION**
**Corresponding Author**

Maolin Bo − Key Laboratory of Extraordinary Bond Engineering and Advance Materials Technology (EBEAM) of Chongqing, School of Materials Science and Engineering, Yangtze Normal University, Chongqing 408100, P. R. China; orcid.org/0000-0003-1173-8919
Phone: 86-13787428880; Email: bmlwd@yznu.edu.cn

**Authors**

Liangjing Ge − Key Laboratory of Extraordinary Bond Engineering and Advance Materials Technology (EBEAM) of Chongqing, School of Materials Science and Engineering, Yangtze Normal University, Chongqing 408100, P. R. China;

Jibiao Li − Key Laboratory of Extraordinary Bond Engineering and Advance Materials Technology (EBEAM) of Chongqing, School of Materials Science and Engineering, Yangtze Normal University, Chongqing 408100, P. R. China;

Lei Li−Key Laboratory of Extraordinary Bond Engineering and Advance Materials Technology (EBEAM) of Chongqing, School of Materials Science and Engineering, Yangtze Normal University, Chongqing 408100, P. R. China;

Chuang Yao−Key Laboratory of Extraordinary Bond Engineering and Advance Materials Technology (EBEAM) of Chongqing, School of Materials Science and Engineering, Yangtze Normal University, Chongqing 408100, P. R. China; orcid.org/0000-0002-6673-8388

Zhongkai Huang−Key Laboratory of Extraordinary Bond Engineering and Advance Materials Technology (EBEAM) of Chongqing, School of Materials Science and Engineering, Yangtze Normal University, Chongqing 408100, P. R. China;

**Figure and Table Captions**

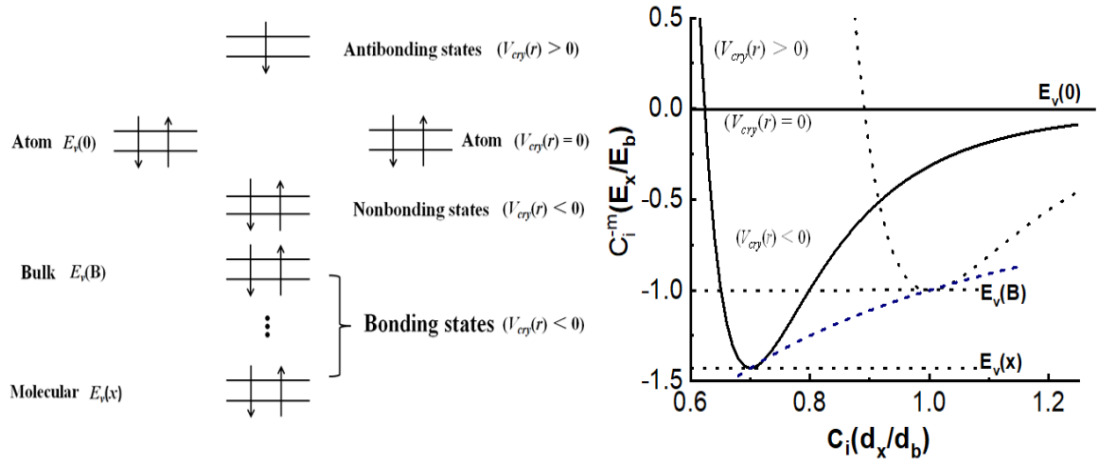

**Fig. 1** Schematic of atomic bonding and binding energy (BB) model in combination with the BOLS notation. The BOLS-BB model obtains atomic bonding by quantifying the binding energy.

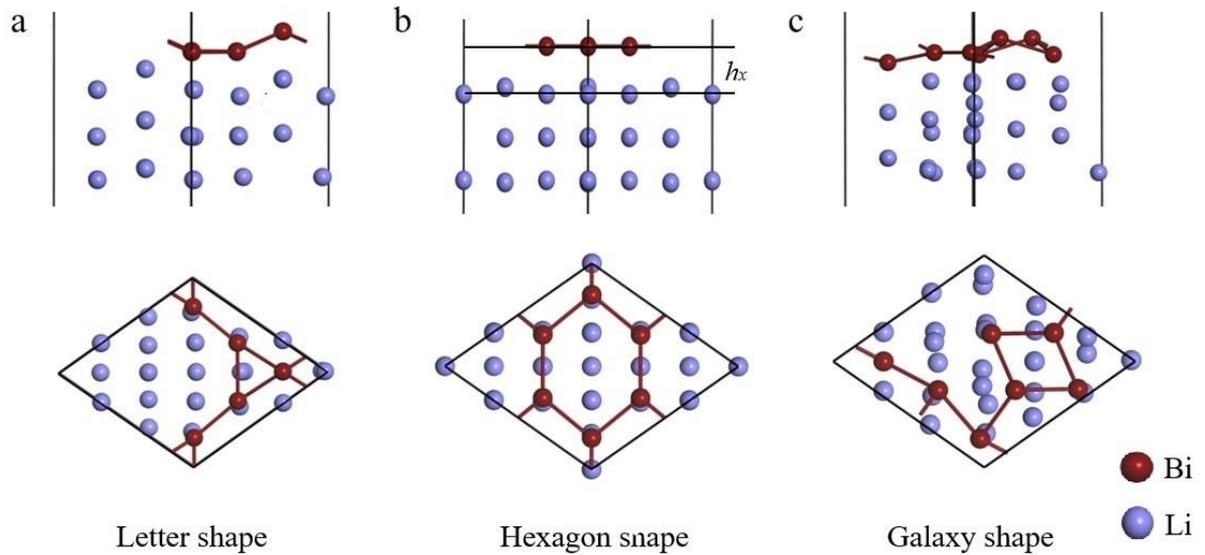

Letter shape    Hexagon shape    Galaxy shape

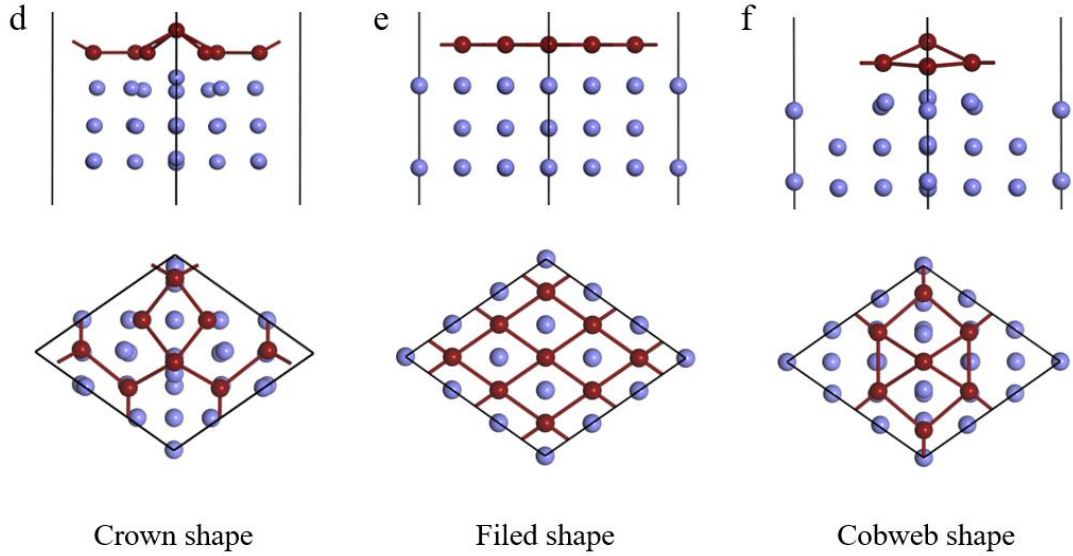

Crown shape          Filed shape          Cobweb shape

**Fig. 2** Optimized geometric structures of Bi atoms adsorbed on Li(110) surface: (a) letter shape, (b) hexagon shape, (c) galaxy shape, (d) crown shape, (e) field shape, and (f) cobweb shape. In each panel, the upper image is a side-on view of the structure and the lower image is a top-down view of the same. The height $h_x$ is shortest distance between the Bi atoms layer and Li(110) atoms layer.

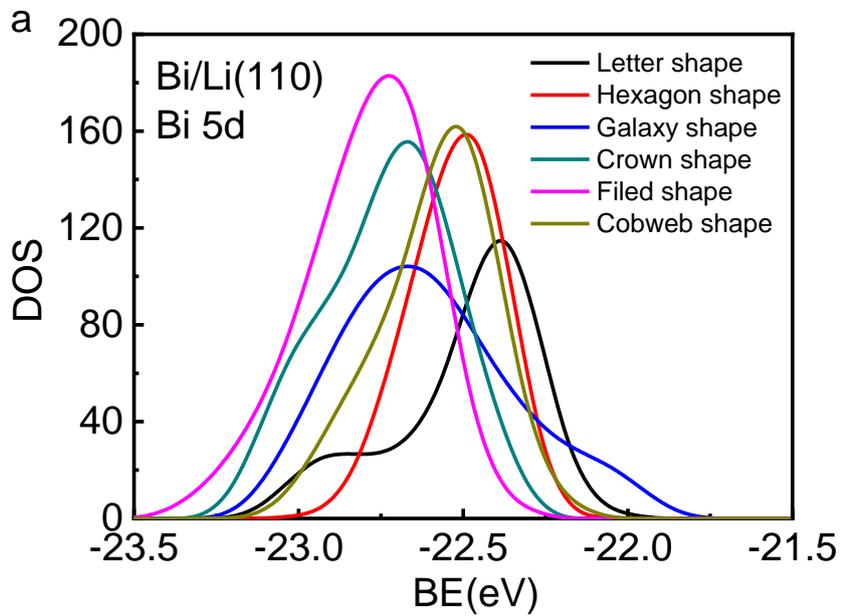

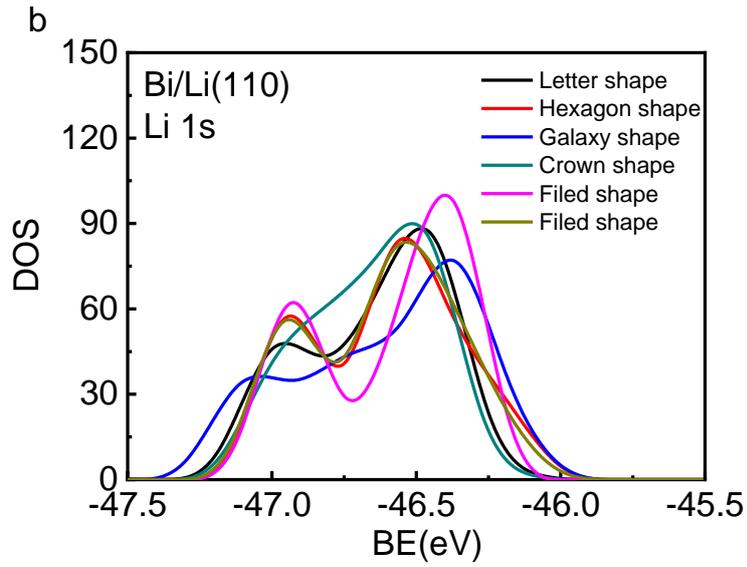

**Fig. 3** (a) Bi $5d$, (b) Li $1s$ DOS for 2D metal Bi/Li(110) letter-, hexagon-, galaxy-, crown-, field-, and cobweb-shaped heterojunction structures.

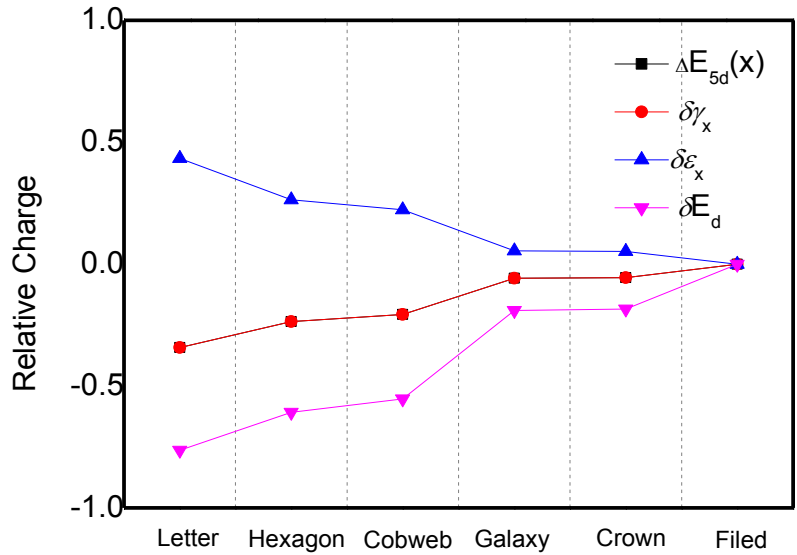

**Fig. 4** Trends for energy shift $\Delta E_{5d}(x)$, RBED $\delta E_d$, RLBS $\delta\varepsilon_x$, and RBER $\delta\gamma_x$ as predicted by the BOLS–BB model and DFT calculations.

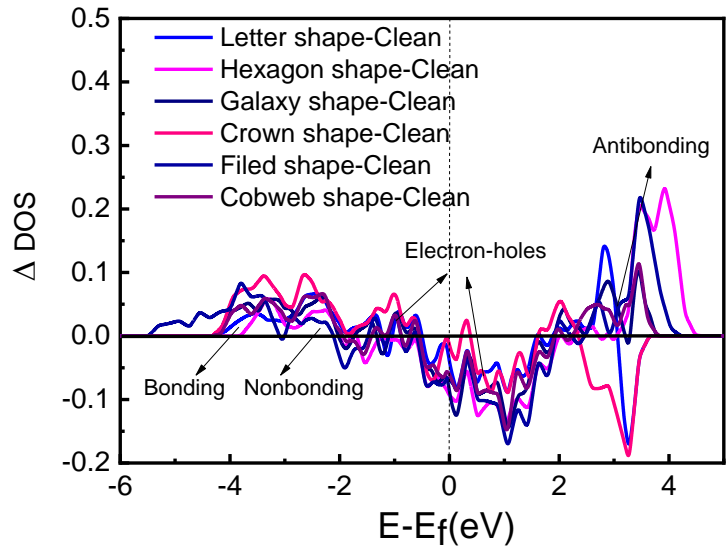

**Fig. 5** Valence band of Bi/Li(110) heterostructures. The profiles exhibit four valence DOS features: antibonding, electron–hole, nonbonding, and bonding states. The dotted line in the figure is the Fermi level.

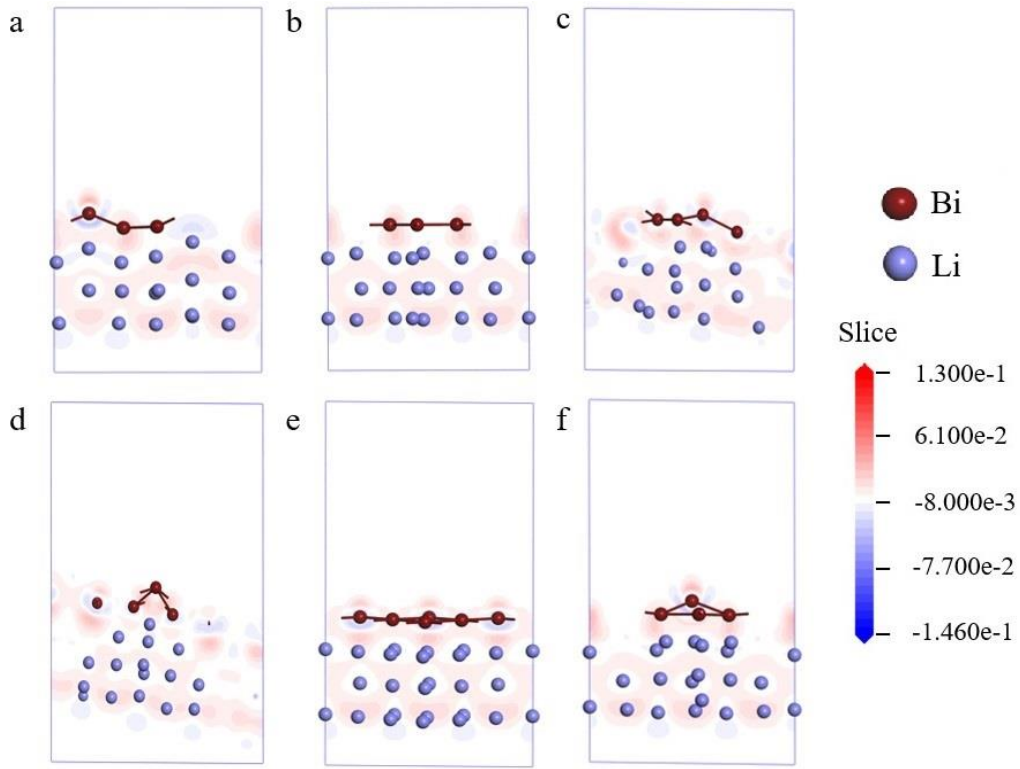

**Fig. 6** Deformation charge density of (a) letter-, (b) hexagon-, (c) galaxy-, (d) crown-, (e) field-, and (f) cobweb-shaped structures of Bi atoms.

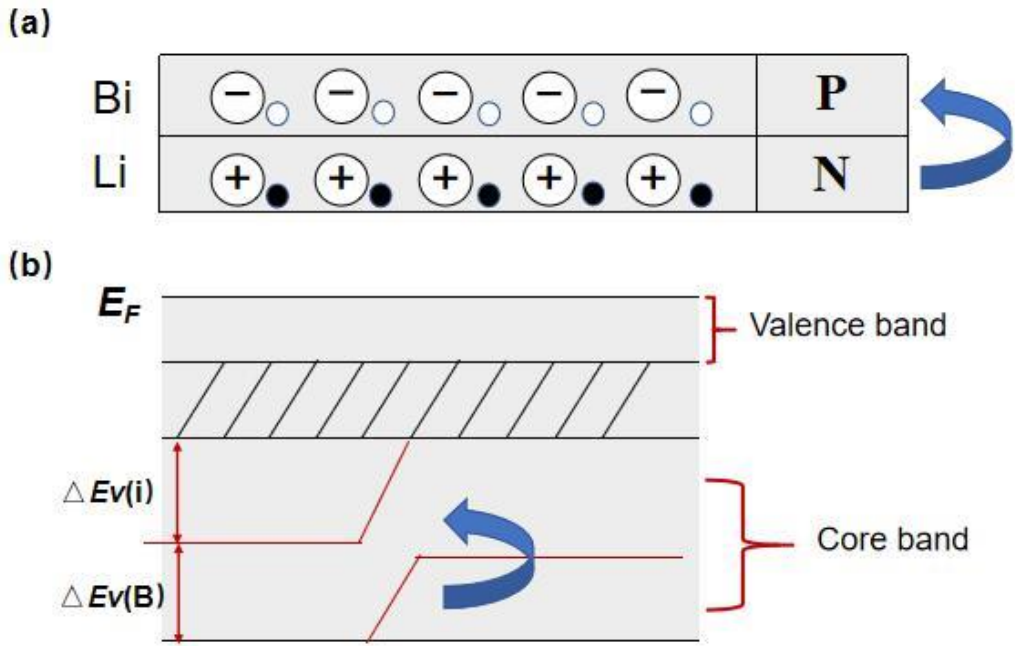

**Fig. 7** Schematics showing (a) 2D metal PN junctions and (b) electronic energy shift of core band structure.

**Table 1** Geometric and energy parameters for different shaped heterojunctions as obtained via first-principles calculations. Table gives the corresponding values of the Work function (eV), the height $h_x$ (Å) and the total energy(eV) of the relaxed Bi/Li(110) heterostructure, isolated Bi, and Li monolayers, corresponding binding energies $_{Eb}$ of the listed each heterostructures. The shortest distance between the Bi atoms adsorbed on the Li(110) surface of height $h_x$ (Å).

| Geometric structures | $h_x$(Å) | Work function (eV) | | | Total Energy (eV) | | | $E_b$ (eV) |
|---|---|---|---|---|---|---|---|---|
| | | Bi | Li | Bi/Li(110) | $E_{Bi}$ | $E_{Li}$ | $E_{Bi/Li}$ | |
| Letter shape | 0.95 | 4.183 | 2.903 | 3.266 | -10630.71 | -5460.24 | -16097.72 | -6.77 |
| Hexagon shape | 2.09 | 4.429 | 3.019 | 3.423 | -12757.54 | -5460.38 | -18225.79 | -7.87 |

| | | | | | | | |
|---|---|---|---|---|---|---|---|
| Galaxy shape | 1.10 | 4.256 | 3.037 | 3.366 | -14883.39 | -5459.5 | -20352.67 | -9.78 |
| Crown shape | 1.59 | 4.188 | 3.016 | 3.255 | -17011.06 | -5460.09 | -22479.13 | -7.98 |
| Field shape | 2.36 | 4.224 | 3.070 | 3.354 | -19135.4 | -5460.42 | -24603.65 | -7.83 |
| Cobweb shape | 1.76 | 4.314 | 3.030 | 3.328 | -14884.11 | -5460.3 | -20352.29 | -7.88 |

**Table 2** The electronic BE of the Bi 5$d$ level is written $E_{5d}(x)$ (eV); the energy shifts are computed as $\Delta E_{5d}(x) = E_{5d}(x) - E_{5d}(B)$ (eV); potential energy of inter-surface $V_{\text{surface}}(r_i) = (\frac{1}{4\pi\varepsilon_0})\frac{q_1(+)q_2(-)}{r_x}$; values of the charges of the Bi and Li atoms; $\gamma_x = \frac{\Delta E_v(B) + \Delta E_v(x)}{\Delta E_v(B)}$ is bond energy density; the relative bond energy ratio (RBER) $\delta\gamma_x = \gamma_x - 1$; the relative bond energy density (RBED) $\delta E_d = \gamma_x^4 - 1$; and relative local bond strain (RLBS) $\delta\varepsilon = \gamma_x^{-1} - 1$ are given.

| Geometric structures | $E_{5d}(x)$ | $\Delta E_{5d}(x)$ | $\delta\gamma_x$ | $\delta\varepsilon_x$ (%) | $\delta E_d$ (%) | Charge[a] (Bi) (e/atom) | Charge[a] (Li) (e/atom) | $V_{\text{surface}}$ |
|---|---|---|---|---|---|---|---|---|
| Letter shape | 22.385 | -0.342 | -0.342 | 0.434 | -0.764 | -0.482 | 0.326 | -4.762 |
| Hexagon shape | 22.491 | -0.236 | -0.236 | 0.264 | -0.608 | -0.490 | 0.405 | -2.734 |
| Galaxy shape | 22.669 | -0.058 | -0.058 | 0.054 | -0.190 | -0.466 | 0.424 | -5.171 |
| Crown shape | 22.671 | -0.056 | -0.056 | 0.052 | -0.184 | -0.307 | 0.368 | -2.046 |
| Field shape | 22.727 | 0 | 0 | 0 | 0 | -0.382 | 0.440 | -2.050 |
| Cobweb shape | 22.521 | -0.206 | -0.206 | 0.223 | -0.553 | -0.417 | 0.397 | -2.708 |

[a] Negative signs indicate charge gain; otherwise, charge loss occurs.